\documentclass[runningheads]{llncs}

\usepackage[T1]{fontenc}
\usepackage{graphicx}
\usepackage{hyperref}
\usepackage{color}

\usepackage[locale=US]{siunitx}
\usepackage[capitalise]{cleveref}
\usepackage{bm}
\usepackage{mathptmx}
\usepackage{listings}

\DeclareSIUnit\angstrom{\text {Å}}

\begin{document}

\title{Load balancing for adaptive-precision interatomic potentials in materials science}

\author{David Immel\inst{1}\orcidID{0000-0001-5143-8043} \and Godehard Sutmann\inst{1,2}\orcidID{0000-0002-9004-604X}}

\authorrunning{D. Immel and G. Sutmann}

\institute{Jülich Supercomputing Centre (JSC), Forschungszentrum Jülich, Jülich, Germany \and Interdisciplinary Centre for Advanced Materials Simulations (ICAMS), Ruhr Universität Bochum, Bochum, Germany\\
\email{d.immel@fz-juelich.de}
}

\maketitle

\begin{abstract}
For atomistic molecular dynamics simulations, we consider a recently developed hybrid coupling between the highly accurate machine learning (ML)-based atomic cluster expansion (ACE) interaction model and a less precise (but about 1-2 orders faster) EAM potential, in order to leverage the performance bottleneck of pure ML potentials in a parallel computing environment.
This spatial-temporal adaptivity has the potential for a speedup of more than an order of magnitude, that would be lost without a dynamic load-balancing strategy that becomes critical due to fluctuating potential contributions.
We compare the load-balancing strategies of LAMMPS with load-balancing methods provided by the library ALL in adaptive-precision EAM-ACE simulations of a copper nanoindentation.
For a regular-grid domain decomposition, the modified tensor method of the ALL library, which conserves the simplicity and transferability of the tensor method, significantly reduces the imbalance factor of the force calculation from 3.0 to 1.7 compared to the native load-balancing method of LAMMPS.
A nearly perfectly balanced system with an imbalance factor close to 1.0 can be reached by load balancing an irregular grid of domains with the recursive bisectioning method of LAMMPS and with ALL's histogram method for a staggered grid of domains.
Load balancing a staggered grid with ALL is found to be significantly faster than load balancing an irregular grid with the bisectioning method of LAMMPS, but the communication costs during a regular timestep are lower for the irregular grid of domains, as a lower surface-by-volume ratio of the domains is reached.
Furthermore, we suggest a strategy to find the ideal load-balancing frequency for adaptive-precision simulations.

\keywords{load balancing \and adaptive-precision interatomic potentials \and spatial domain decomposition \and molecular dynamics \and parallel computing}
\end{abstract}

\section{Introduction}
\label{sec:introduction}
Modern machine-learning potentials remain orders of magnitudes slower than traditional empirical potentials but offer the highest accuracy available~\cite{pace,doi:10.1021/acs.jctc.2c01149,Stark_2024}.
Thus, the coupling of a fast surrogate model with an accurate but computationally less efficient interatomic potential can save computational time~\cite{adaptive_precision_potentials,paper_nanoindentation,conservative_paper,Birks2026}.
Another motivation for the coupling of two potentials is the simulation of two different states (ground and excited) which cannot be described by one potential~\cite{Thompson2026}.
Therefore, the development of coupling approaches is currently an active field of research~\cite{adaptive_precision_potentials,paper_nanoindentation,conservative_paper,Birks2026,Thompson2026}.

All these approaches use a switching parameter $\lambda_i$ that interpolates between the fast and the accurate model per atom $i$, i.e.,
\begin{equation}
\gamma_i = \lambda_i \gamma_i^\text{fast} + (1-\lambda_i) \gamma_i^\text{accurate}\,,
\label{eq:gamma}
\end{equation}
where the interpolated property $\gamma_i$ can be the potential energy~\cite{adaptive_precision_potentials,conservative_paper} or atomic force~\cite{Thompson2026,Birks2026}.
When $\lambda_i=1$ applies for the majority of the atoms, the computation of the accurate and computationally expensive $\gamma_i^\text{accurate}$ is only required for a subset of atoms.
Therefore, one can achieve -- dependent on the coupled models and atomistic system -- a speedup of 1-2 orders of magnitude compared to a purely accurate calculation~\cite{adaptive_precision_potentials,paper_nanoindentation,conservative_paper,Birks2026}.

As one can save wall-time and calculate larger systems with parallel computation, atomistic simulations account for a relevant share of applications on today's supercomputers and high-performance computing clusters~\cite{jupiter,anton3,noctua2}.
Therefore, one needs to analyze the requirements of the novel adaptive-precision models in terms of parallel computing.
In parallel simulations, the simulation box is divided into disjoint domains which cover the whole simulation box according to a spatial domain-decomposition approach\cite{spatial_domain_decomposition_lammps,spatial_domain_decomposition_gromacs,spatial_domain_decomposition_amber}.
Each MPI task, in the following denoted as processor, is assigned to one domain and administers all particles located within this domain.
Such a domain decomposition is a natural and beneficial approach for molecular dynamics (MD) as (most) interactions are local and depend only on neighboring particles within a short cutoff-radius.
Thus, each processor sends the position of particles located near its domain boundary to neighboring processors prior to the force-computation and receives the forces on these particles back after the force-computation~\cite{BOWERS2007303,midpoint_method}.
The communication of positions and forces prior and after the force calculation synchronizes all processors.
Therefore, the computational time required for the force calculation -- the most time-consuming part of a MD simulation~\cite{spatial_domain_decomposition_amber} -- should be equal for all processors to prevent idle times.
Therefore, we define the imbalance $I$ as~\cite[p. 1391]{lammps_manual}
\begin{equation}
I = \frac{\text{max}_p(\tau^\text{F}_p)}{\langle\tau^\text{F}_p\rangle_p},
\label{eq:imbalance}
\end{equation}
where $\tau^\text{F}_p$ denotes the force-calculation time measured on processor $p$ and $I=1$ corresponds to a perfectly balanced system.

There are two typical causes for load imbalances in static systems calculated with constant-precision potentials.
Firstly, the particles may be unevenly distributed among the processors.
Secondly, the interaction density may be inhomogeneous, caused by an inhomogeneous particle distribution or surfaces at box boundaries (cf. Ref. ~\cite{adaptive_precision_potentials}).
For dynamic systems, particles moving between domains can cause load imbalance, in particular for domains with few particles.
Therefore, particle simulations may use initial static load balancing or dynamic load balancing during a simulation to prevent these load-imbalances by adjusting the spatial volume administered by each processor.

Adaptive-precision simulations introduce with the choice of the used model per particle and timestep another load-imbalance source: Either the fast model ($\lambda_i=1$), the accurate model ($\lambda_i=0$) or both models are required ($\lambda_i\in(0,1)$).
Thus, while dynamic load-balancing may be considered optional for constant-precision simulations, it is essential for parallel adaptive-precision simulations.

\begin{figure}[tb]
\centering
\includegraphics[width=0.20\columnwidth]{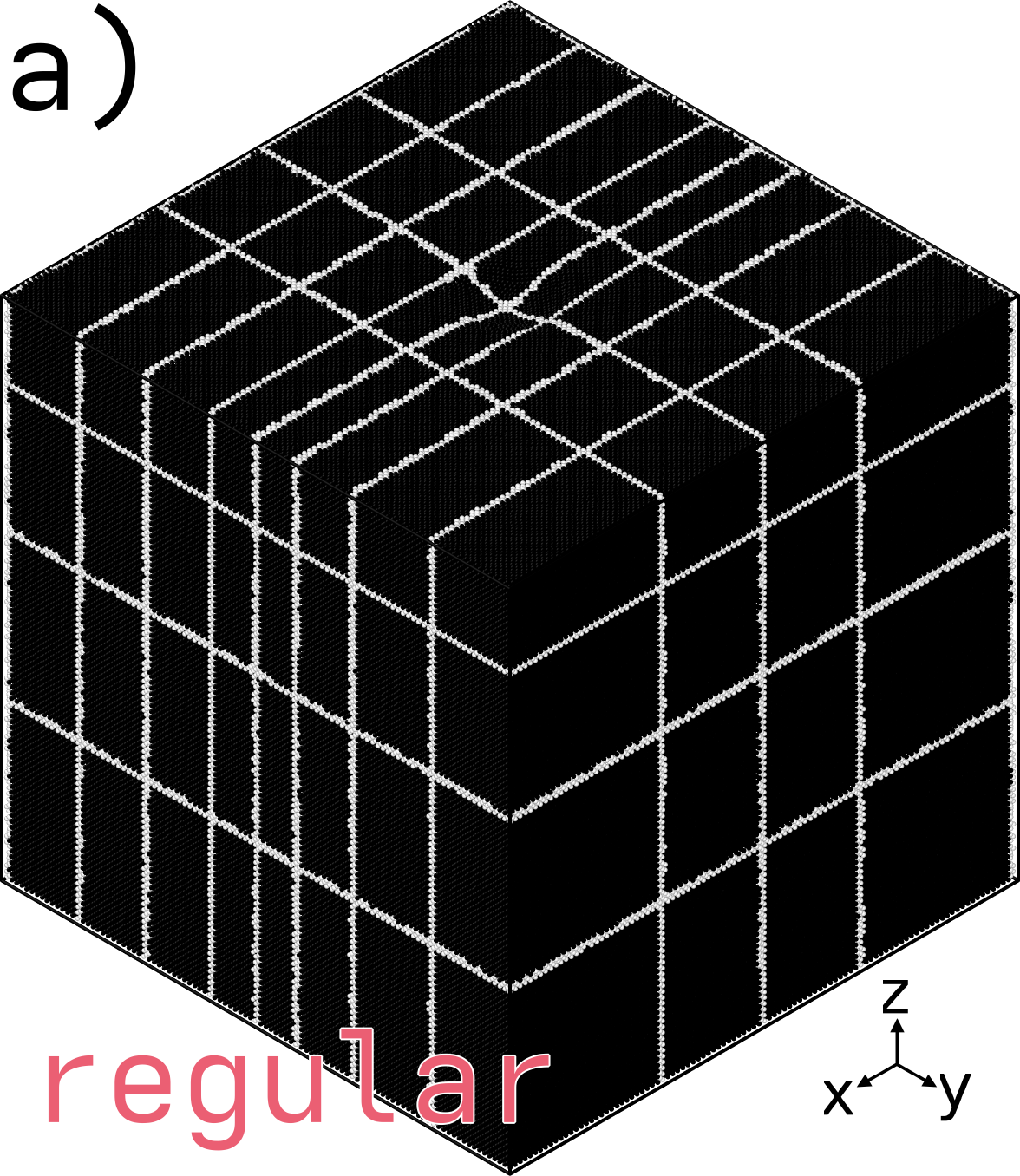}
\includegraphics[width=0.20\columnwidth]{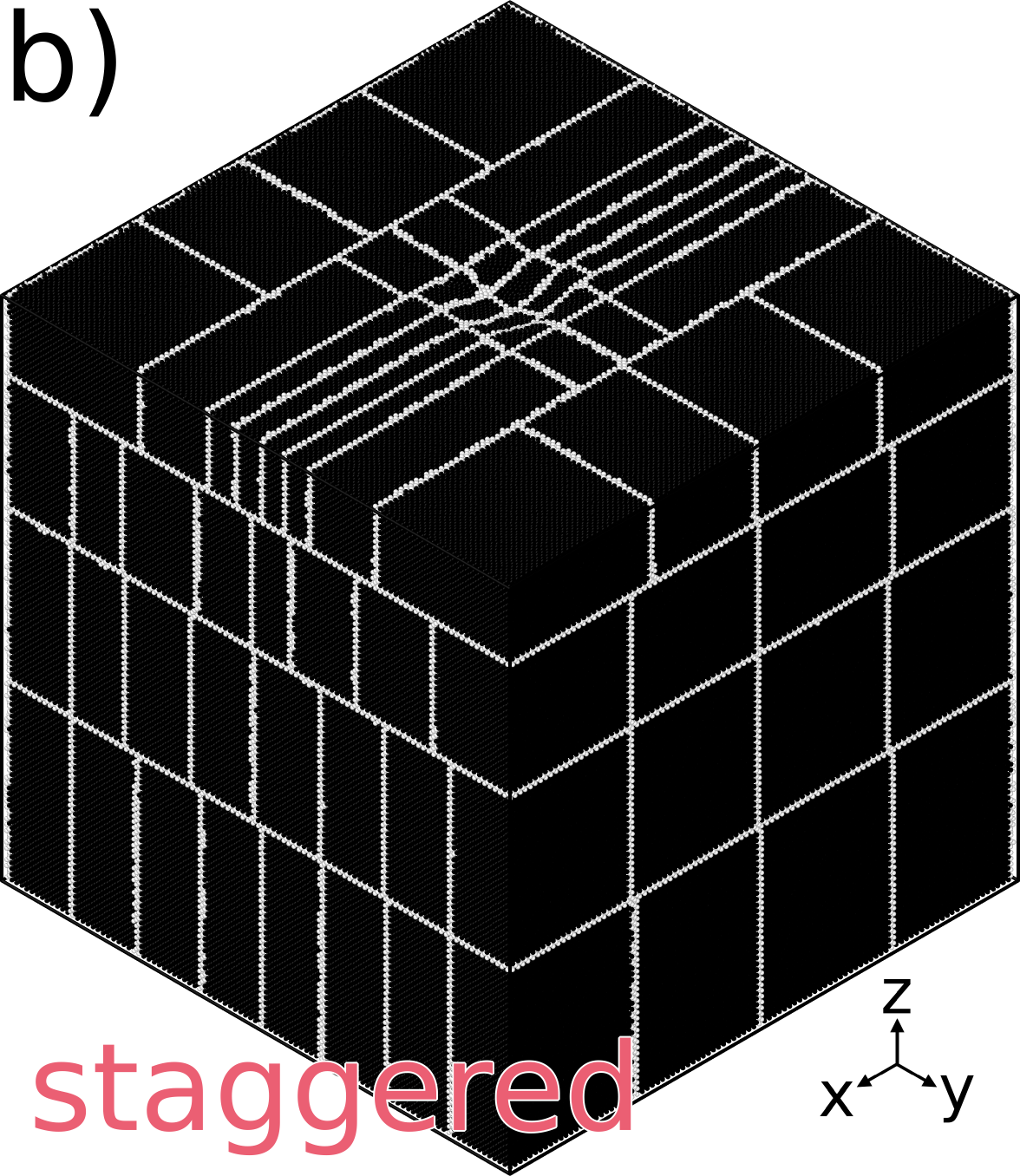}
\includegraphics[width=0.20\columnwidth]{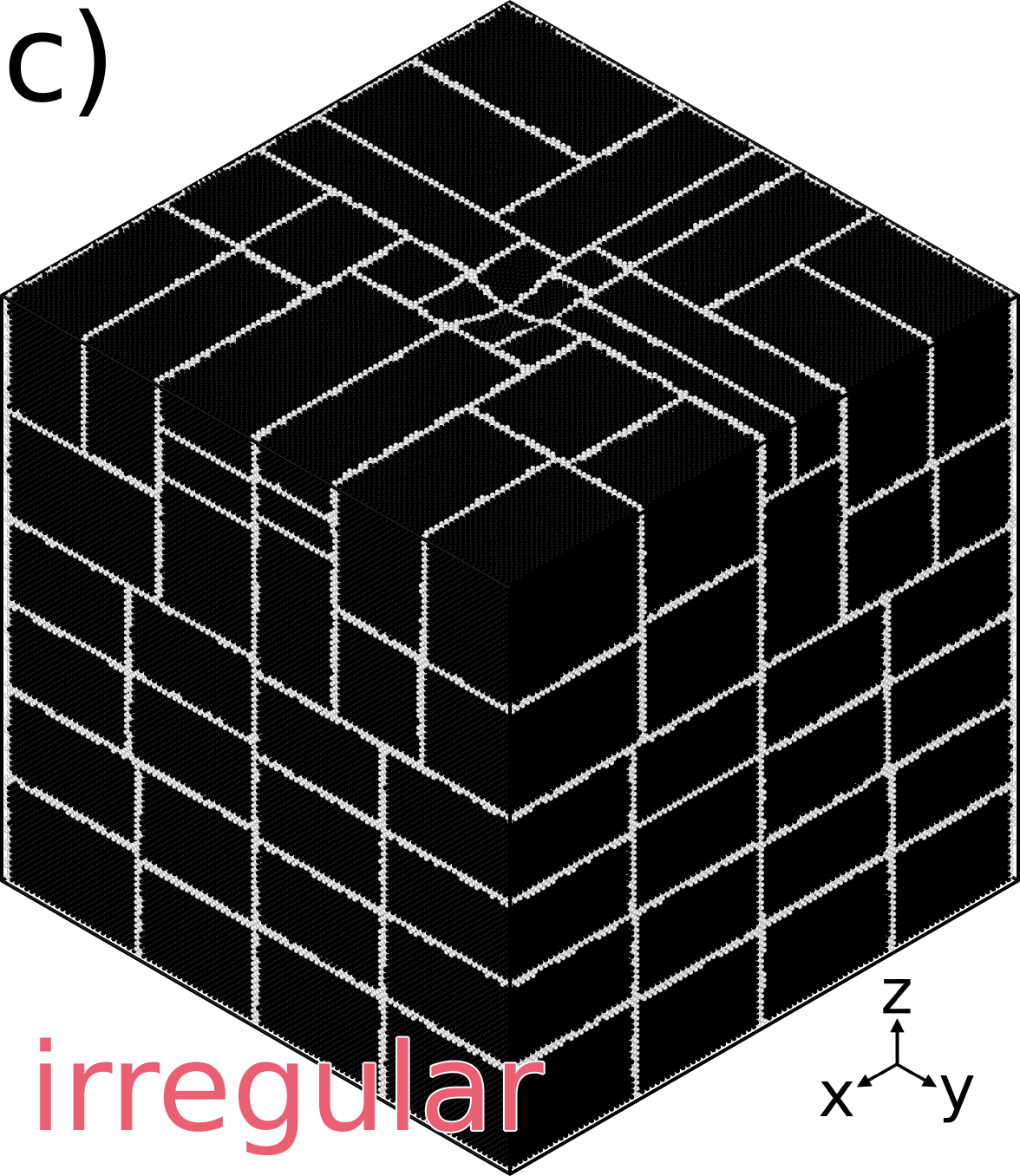}
\caption{\label{fig:grid:types}
Spatial domain decomposition using a a) regular, b) staggered, c) irregular grid.
}
\end{figure}

There are different domain decompositions.
The simplest domains are rectangular cuboids~\cite{7839674,lammps}, but there are also more complex forms~\cite{BEGAU201551,FATTEBERT20122608,spatial_domain_decomposition_gromacs}.
Rectangular cuboids are widely used as domains~\cite{lammps,GLASER201597,JackyKoerner2021,rene_staggered_conference_paper,ls1mardyn} and can be arranged in a regular grid~\cite{lammps,GLASER201597,JackyKoerner2021}, a staggered grid~\cite{rene_staggered_conference_paper,spatial_domain_decomposition_gromacs,Seckler2021} or in an irregular grid~\cite{lammps,ls1mardyn}.
Examples of these three grid types are visualized in \cref{fig:grid:types} with OVITO~\cite{ovito}.
Regular grids are the easiest to implement, allow the optimization of communication as there are only 26 neighbors per domain~\cite{Zhao2023}, but may fail to balance the load due to constrained domain boundaries~\cite{rene_staggered_conference_paper}.
Staggered grids are less constrained and, thus, allow to efficiently adapt to load distributions~\cite{rene_staggered_conference_paper}.
Furthermore, a staggered grid can be dynamically balanced with local information~\cite{ALL}.
Irregular grids can also efficiently adapt to load distributions and be created by recursive bisections under consideration of the surface-to-volume ratio at every bisection~\cite{https://doi.org/10.1002/nme.2184}.
The surface of the domains should be minimal to minimize the communication required prior and after the force calculation.
Therefore, irregular grids can minimize the number of communicated atoms although the communication patterns become more complex and expensive compared to a regular grid~\cite{Zhao2023}.

The Large-scale Atomic/Molecular Massively Parallel Simulator (LAMMPS)~\cite{lammps} is a modularly designed open-source MD code, that allows the usage of adaptive-precision potentials~\cite{adaptive_precision_potentials,Birks2026,paper_nanoindentation,conservative_paper,Thompson2026}.
LAMMPS provides dynamic load-balancing via recursive coordinate bisectioning\cite{berger1987rcb} for irregular grids and domain-boundary shifts for a regular grid.
The staggered-grid decomposition for LAMMPS is introduced with the present work.
Furthermore, we compare multiple load-balancing strategies for regular and staggered grids provided by the ALL-library~\cite{ALL} with existing load-balancing routines of LAMMPS.

\section{Methods}
\label{sec:methods}
\subsection{Spatial domain decompositions}
\label{sec:spatial:domain:decompositions}
A domain decomposition using a regular grid (cf. \cref{fig:grid:types}a) consists of $P=P_\text{x} P_\text{y} P_\text{z}$ domains with $P_\text{x}+1$ domain boundaries $x_i$ in the x-dimension and $P_\text{y}+1$ and $P_\text{z}+1$ domain boundaries $y_i$ and $z_i$ in the y- and z-dimension.
The boundaries at the simulation box are usually fixed~\cite{ALL,lammps}, thus a such a regular grid has $P_\text{x}+P_\text{y}+P_\text{z}-3$ degrees of freedom.
There are different methods to achieve a better balanced regular grid.
The balancing of load between two domain boundaries is implemented in LAMMPS~\cite{lammps_balance_command}. As that the load between two domain boundaries is distributed about multiple processors according to the boundaries in the other two spatial dimensions, e.g., $P_\text{y}P_\text{z}$ processors when the boundaries $\{x_i\}$ are balanced, a perfect balance between all domains cannot be expected for all systems.
Alternatively, one can try to minimize the maximal load of a processor between domain boundaries as implemented in ALL~\cite{ALL}.

A staggered grid (cf. \cref{fig:grid:types}b) is created by recursively cutting the simulation box into different layers, each layer into different rows, and each row into the domains.
As all these cuts are independent of each other, there are $P_\text{z}-1$ layer-cuts, $P_\text{z}(P_\text{y}-1)$ row cuts and $P_\text{z}P_\text{y}(P_\text{x}-1)$ cell cuts; in total $P-1$ cuts.
Thus, a staggered grid can in theory -- in contrast to a regular grid -- perfectly adapt to the load distribution.
In practice, this might not be possible for load distributions that are not continuous enough like a regular grid of particles at 0K.

An irregular grid is usually generated by recursive coordinate bisections (RCB)~\cite{berger1987rcb}, that recursively divide the simulation box into subvolumes with equal load until there is one subvolume per processor.
In contrast to the shifted domains of a staggered or regular grid, the cuts that generate the irregular domains are not shifted by a load balancer but done completely new always starting from the first cut~\cite{7839674}.
Therefore, global communication during load-balancing of a RCB is unavoidable.
A staggered grid can be created by RCB when all cuts between two subdomains in the first iterations of the recursion are in one dimension, then all cuts in the next iterations of the recursion are in a second dimension and all remaining iterations cut subdomains only in the third dimension.
Therefore, an RCB-domain decomposition has the same number of independently adjustable domain boundaries like a staggered grid, i.e., $P-1$.

\subsection{Load-balancing methods}
\label{sec:load:balancing:methods}
There are two different load-balancing methods implemented in LAMMPS~\cite[p. 1389]{lammps_manual}.
\begin{itemize}
\item \emph{LMP shift} aims at balancing the load between cuts of a regular grid based on an atomic load. The new cuts are searched using an iterative method. The maximum number of iteration is adjustable. We used 10.~\cite[p. 1393]{lammps_manual}.
\item \emph{LMP bisection} aims at balancing the load of the processors through recursively applied bisections based on an atomic load.
\item \emph{LMP report} does not apply load-balancing but reports the imbalance and is ideal as reference simulation.
\end{itemize}

The load-balancing library ALL~\cite{ALL} provides several load-balancing methods for regular grids and staggered grids.
ALL uses the current domain decomposition and the total load or a load histogram per domain as input.
A new domain decomposition is calculated assuming that the load is distributed homogeneously in every domain.
This assumption of homogeneity is obviously incorrect for adaptive-precision simulations.
Nevertheless, the comparison with the native LAMMPS load-balancing methods is promising as ALL offers load-balancing methods that are not used by LAMMPS.
We've implemented several of ALL's load balancing methods for LAMMPS:
\begin{itemize}
\item \emph{ALL tensor classic} aims at balancing the load between cuts of a regular grid based on the total load.
\item \emph{ALL tensor max} aims at decreasing the maximum load between cuts of a regular grid based on the total load.
\item \emph{ALL histogram} shifts the domain boundaries of a staggered grid based on a one-dimensional histogram of the load per domain. The bin width of this histograms can be chosen by the user, $\SI{0.001}{\angstrom}$ in this case as $\SI{0.01}{\angstrom}$ turned out to be too large.
Therefore, $l_\text{b}=\SI{0.001}{\angstrom}$ is used as bin width in \cref{sec:results}.
\item \emph{ALL staggered} shifts the domain boundaries of a staggered grid based on the total load. This method avoids the usage of expensive global communication.
\item \emph{ALL staggered+histogram} uses the expensive ALL histogram method for imbalances larger than a given imbalance threshold. The faster ALL staggered method is used for imbalances smaller than the given imbalance threshold.
\end{itemize}

\subsection{Nanoindentation}
\label{sec:nanoindentation}
\begin{figure}[tb]
\includegraphics[width=.26\columnwidth]{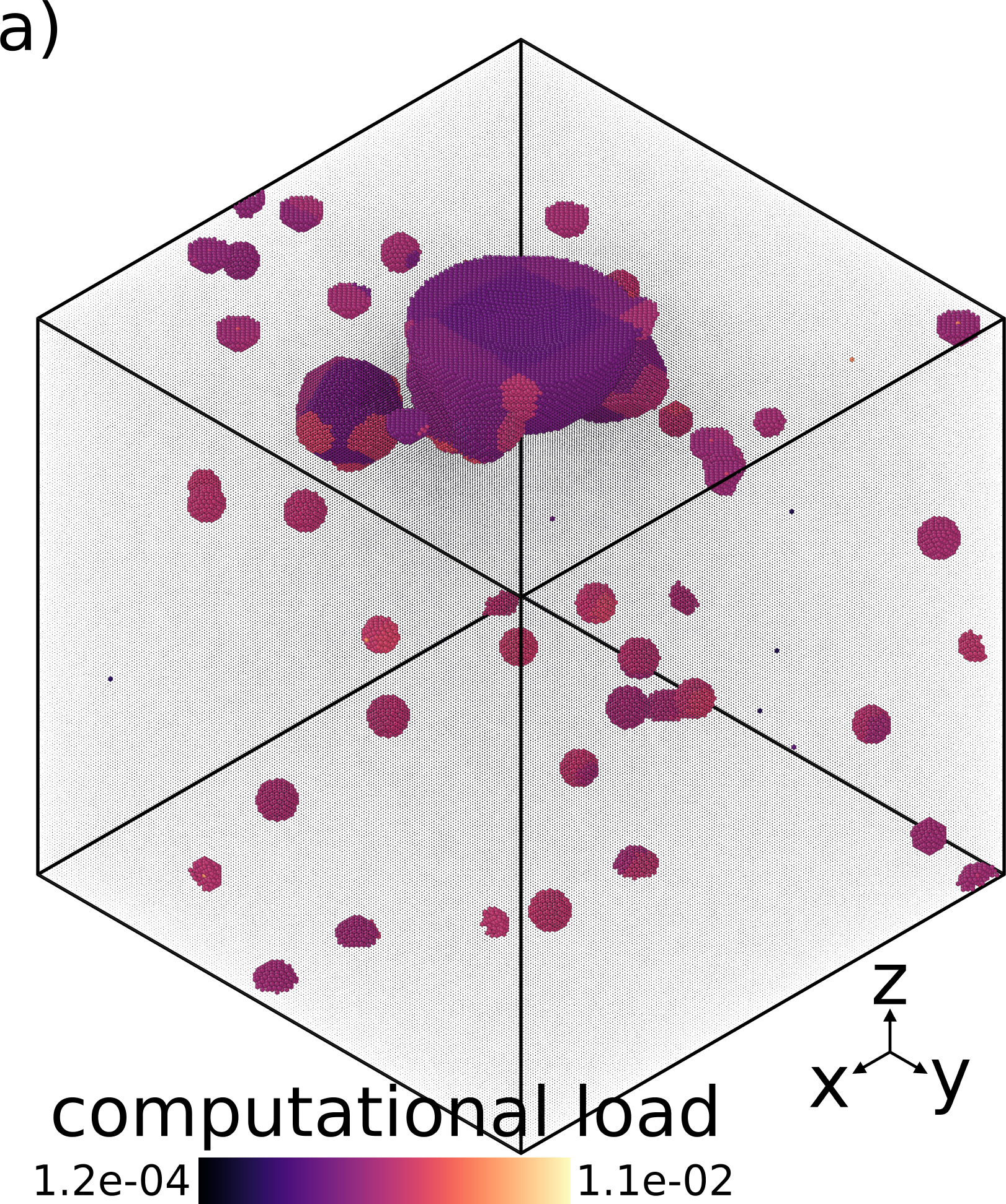}
\includegraphics[width=.65\columnwidth]{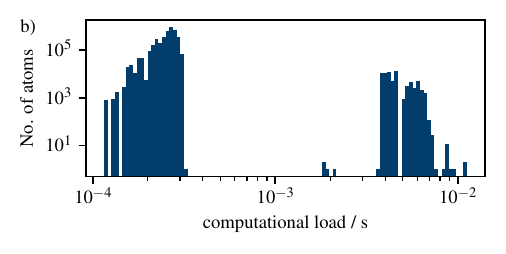}
\caption{\label{fig:nanoindentation:setup}
a) Nanoindentation system after 1000 simulated timesteps.
The atoms are color-coded dependent on their atomic load.
Atoms with a load smaller than $\SI{0.4}{\milli\second}$ are drawn smaller than the remaining atoms with larger loads.
b) Histogram of the atomic load of the system shown in a).
}
\end{figure}
We use a snapshot of the adaptive-precision copper (Cu) nanoindentation from Ref.~\cite{paper_nanoindentation} as starting point for our load-balancing study.
The snapshot was taken after $\SI{60}{\pico\second}$ of the $\SI{100}{\pico\second}$ simulation and contains dislocations.

The adaptive-precision potential couples the potential energies of an atomic cluster expansion (ACE)~\cite{ace} potential and an embedded atom model (EAM)~\cite{DAW1993251} potential according to \cref{eq:gamma}, where the fast EAM Cu potential was adjusted to the precise ACE potential as described in Ref.~\cite{adaptive_precision_potentials}.
We use the same potential and indenter as in Ref.~\cite{paper_nanoindentation}.
Therefore, our load-balancing study uses a realistic test system.

We continue the simulation for 500 or 1000 timesteps ($\SI{0.5}{\pico\second}$ or $\SI{1}{\pico\second}$) to analyze the behavior of ALL's and LAMMPS' load-balancing methods.

The adaptive-precision potential uses a time average of 110 timesteps of the centro-symmetry parameter (CSP)~\cite{csp} to calculate the switching parameter $\lambda_i$.
The time-average was introduced to avoid the usage of the precise ACE potential for thermally fluctuating atoms~\cite{adaptive_precision_potentials}.
The time-averaged CSP was not available to restart the simulation.
Hence, our test simulation is quite dynamic at the start as the average contains only few timesteps and fails in avoiding unintended ACE calculations but becomes better over time.

The load is approximated per atom based on time measurements during the computation of EAM, ACE and the switching-parameter calculation as described in Ref.~\cite{adaptive_precision_potentials}.
The histogram of the atomic load after 1000 timesteps of our test simulation (cf. \cref{fig:nanoindentation:setup}b) shows that 98.2\% of the system have a small computational load ($<\SI{0.33}{\milli\second}$).
The small computational load is on average $\SI{0.26}{\milli\second}$, while the computations of the remaining 1.8\% of the atoms takes on average 18 times longer per atom.
The 3D-visualization in \cref{fig:nanoindentation:setup}a shows that these atoms are distributed over about 40 clusters of varying sizes that are distributed over the whole system.
Therefore, load balancing of adaptive-precision simulations is more challenging than load balancing of constant precision simulations.

The load balancer is called every 10 steps to ensure enough calls during the simulation to allow a meaningful comparison between different methods.
Every 10 timesteps is a high call frequency compared to the intuition gained from constant precision simulations.
However, a load-balancing test performed in Ref.~\cite{adaptive_precision_potentials} has shown that load-balancing every 25 timesteps results in a better balanced system and decreases the required computational resources compared to load-balancing every 100 timesteps.
In this context, load-balancing every 10 timesteps seems a reasonable frequency.
However, we analyze the cost and gain of load-balancing dependent on the load-balancing frequency for our simulations to get a better understanding of the ideal load-balancing frequency.

\section{Results}
\label{sec:results}
\subsection{Regular grid of domains}
\label{sec:results:tensor}

\begin{figure}[p]
\includegraphics[width=2.9in]{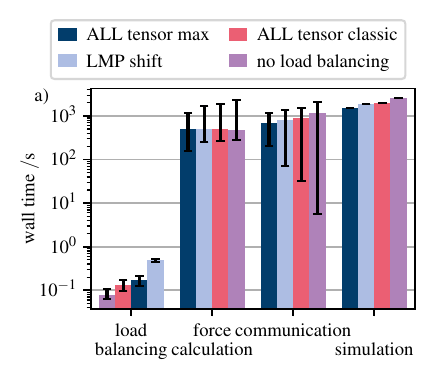}
\includegraphics[width=1.8in]{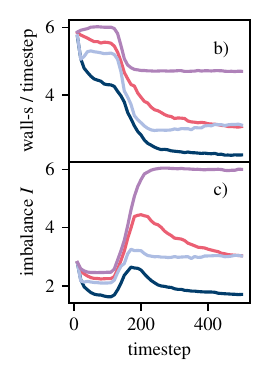}
\includegraphics[width=\textwidth]{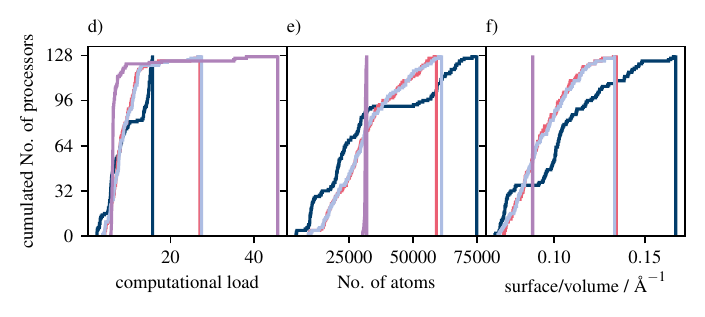}
\includegraphics[width=0.24\textwidth]{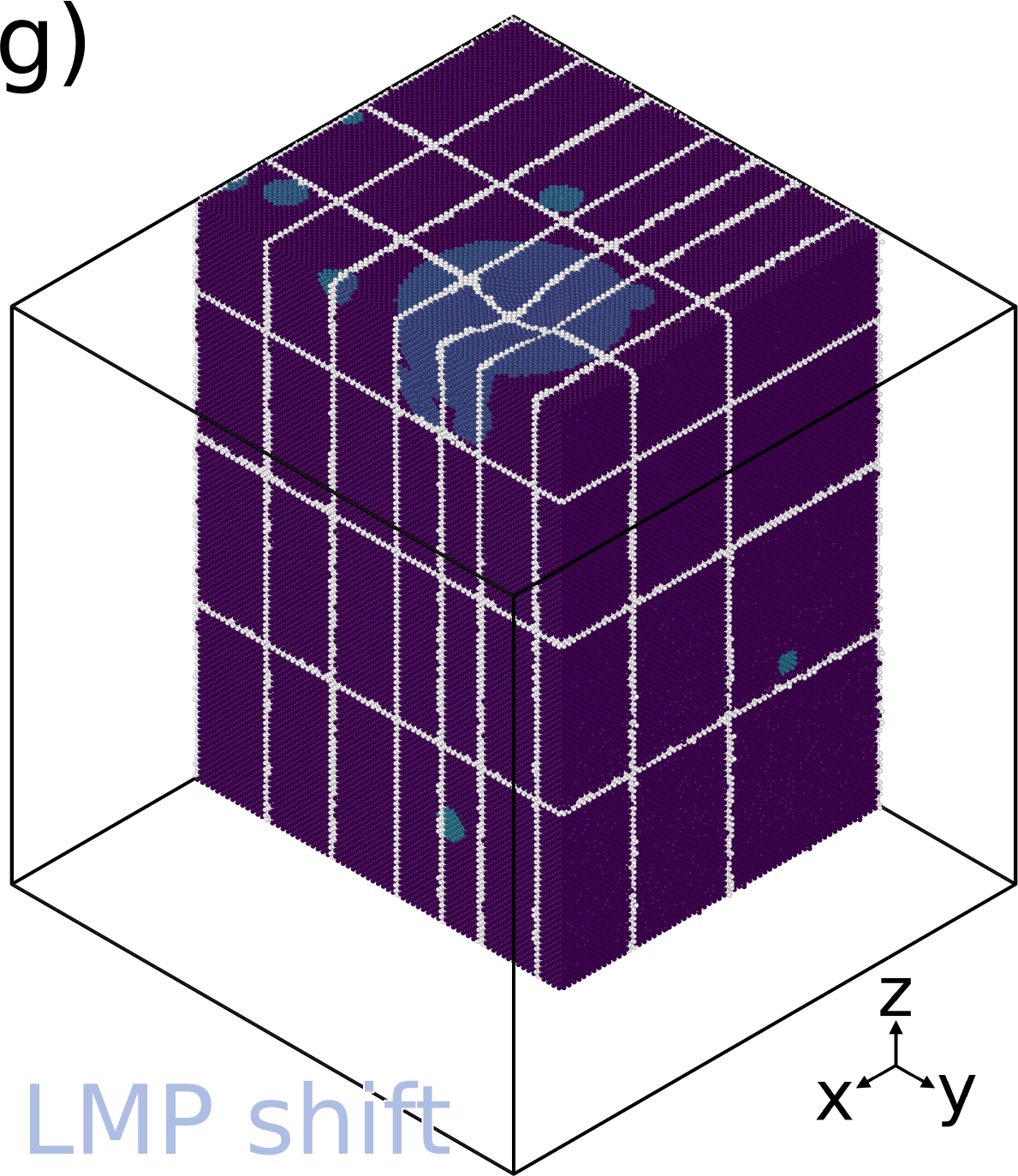}
\includegraphics[width=0.24\textwidth]{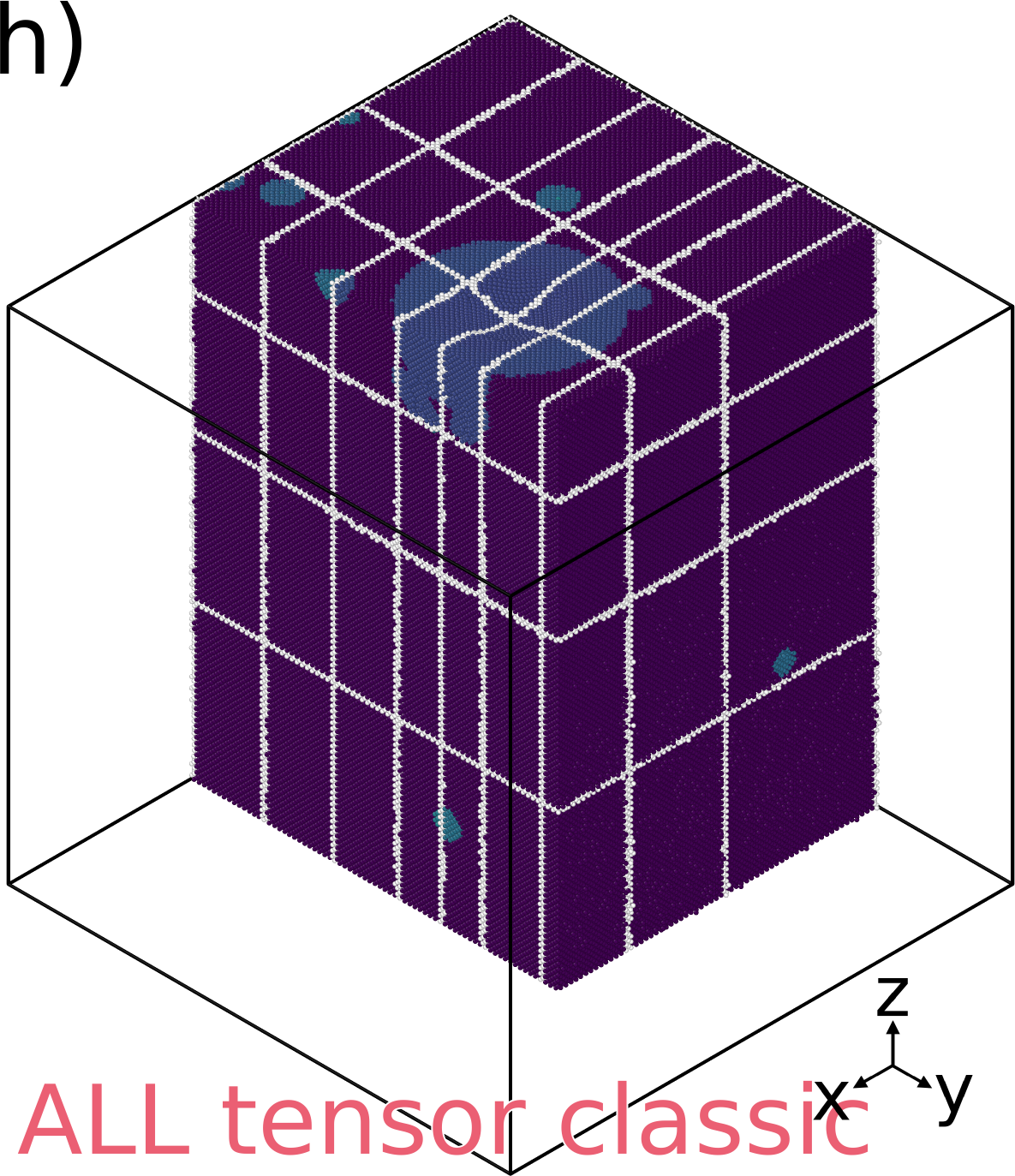}
\includegraphics[width=0.24\textwidth]{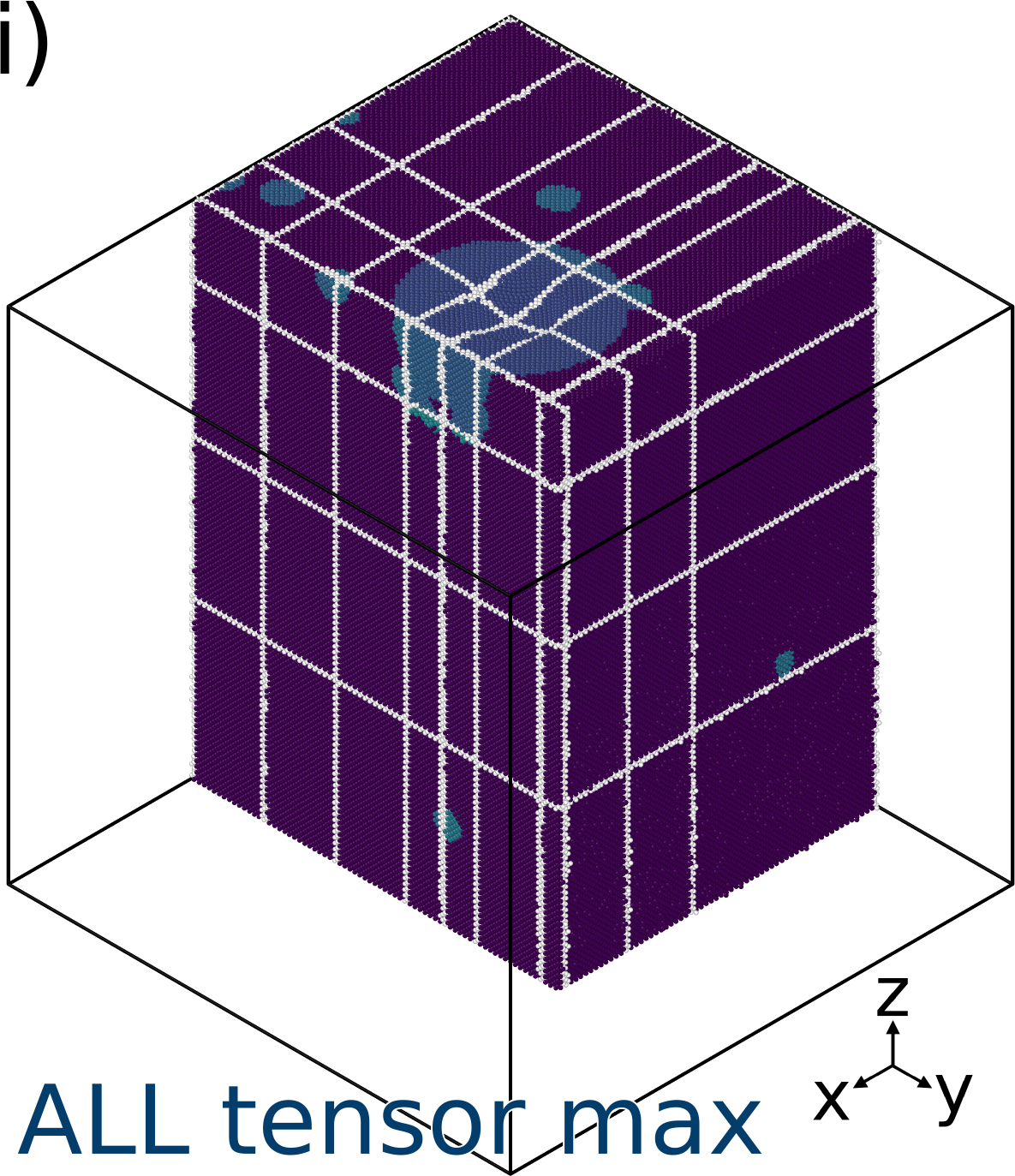}
\includegraphics[width=0.24\textwidth]{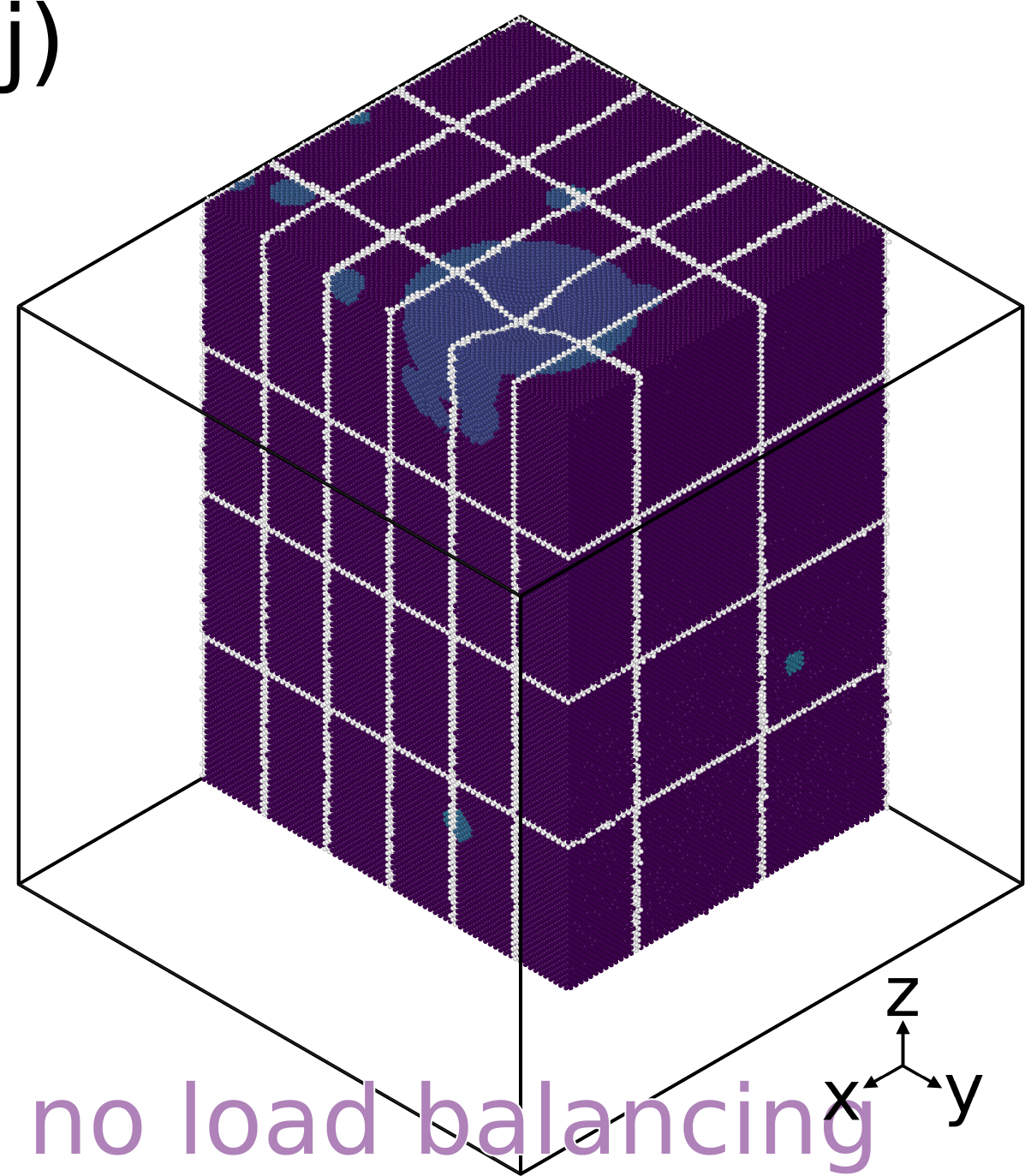}
\caption{\label{fig:tensor}
Analysis of the load-balancing methods for regular grids including
a) the total time spent in some parts of a simulation;
b) the wall-time and c) imbalance dependent on the timestep;
cumulated histograms measured at the end of the simulations of the distribution of the d) computational load, e) number of atoms and f) surface-by-volume ratio among the processors;
and g-j) visualizations of the domain decomposition after the final timestep of the simulations.
}
\end{figure}

We compare three load balancing methods for regular grids in $\SI{0.5}{\pico\second}$-long simulations of the previously introduced nanoindentation system: the ``tensor max'' and the ``tensor classic'' methods of ALL and the native ``shift'' method of LAMMPS.
Plots regarding the analysis of the adaptive-precision simulations are shown in \cref{fig:tensor}.

\Cref{fig:tensor}a shows the fraction of wall time spent for the whole simulation, communication, force calculation and the load balancing.
On average 45\% of the total simulation time (setup time excluded) is spent for communication, but the amount strongly differs between the processors: The minimum and maximum time vary by a factor between 6 (``tensor max'') and 360 (``shift''), indicating that the measured communication time includes waiting times arising from load imbalances.
The average force-calculation time is similar for all balancing methods but not equal as there are double calculations to enable load balancing~\cite{adaptive_precision_potentials}.
Hence, the total force-calculation time depends on the spatial domain decomposition.
The maximum force-calculation time of a processor strongly depends on the load balancing method: The maximum force-calculation time (that causes neighboring processors to wait) is minimal for ``tensor max'' and 1.38 to 1.97 times larger for the other load-balancing methods.

Measuring the spent time dependent on the timestep (cf. \cref{fig:tensor}b) shows that the required time is maximal at the beginning of the simulation and decays during the simulation.
The decaying wall time per timestep is partially caused by the decreasing total load as false positives ($\lambda=0$) are avoided increasingly better by the time average (cf. \cref{sec:nanoindentation}).

Another wall-time reducing factor is load-balancing: The imbalance $I$ (cf. \cref{eq:imbalance}) is decreased from initially about 2.8, becomes worse due to the suppression of precise calculations by the time averaging after about 120 timesteps and is decreased by the load balancer again as shown in \cref{fig:tensor}c.
The final imbalance of ``tensor classic'' and ``shift'' is about the same (3.03) as both methods aim at balancing the load between cuts of the regular grid.
However, the ``shift'' method converges faster than the ``tensor classic'' method, as the actual load of a particle is unknown to the ALL library that provides the new domain boundaries of ``tensor classic'', but known to and used by the ``shift'' method to improve the convergence.
Although the atomic load is also unknown to the ``tensor max'' method, it reaches with 1.7 a far lower imbalance at the end of the simulation that the other methods.
An imbalance of $1$ cannot be reached as the regular grid that defines the domain boundaries has less degrees of freedom than processors as discussed in \cref{sec:spatial:domain:decompositions}.
This example shows that decreasing the maximum load of a processor between cuts of the regular grid is beneficial compared to balancing the total load between cuts ignoring the load that processors actually have.

The cumulated histogram of the load per processor after the final timestep of the simulations is shown in \cref{fig:tensor}d and verifies that ``tensor max'' decreases the maximum load of a processor compared to ``tensor classic'' and ``shift''.
Thereby, the total load between cuts may be different and the total number of particles is more imbalanced as well (cf. \cref{fig:tensor}e,g-j).
Furthermore, the surface-by-volume ratio becomes worse and causes more communication between processors (cf. \cref{fig:tensor}f).
Nevertheless, the total communication cost of the ``tensor max'' domain system is lower than for the other load-balancing methods (cf. \cref{fig:tensor}a) as the idle time is reduced by a larger amount.

\subsection{Staggered or irregular grid of domains}
\label{sec:results:nontensor}

\begin{figure*}[p]
\includegraphics[width=2.9in]{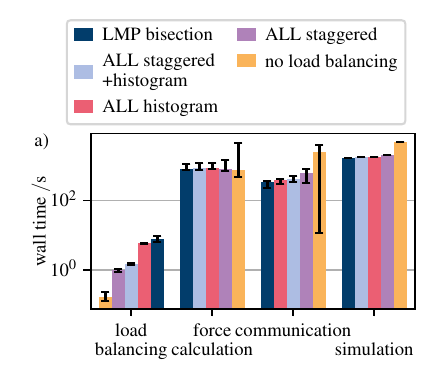}
\includegraphics[width=1.8in]{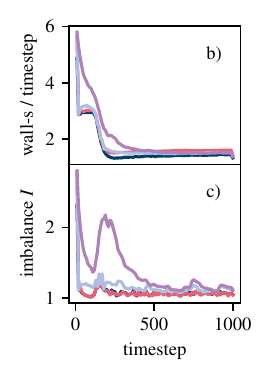}
\includegraphics[width=\textwidth]{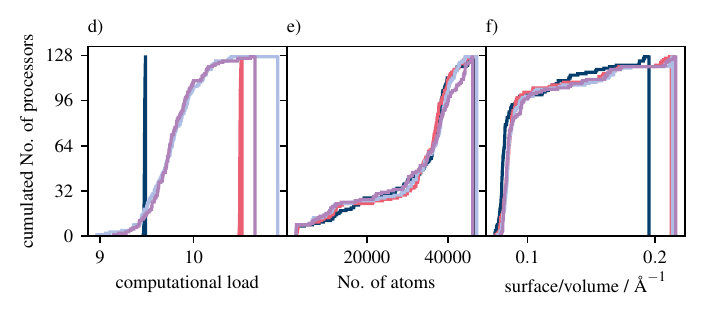}
\includegraphics[width=0.24\textwidth]{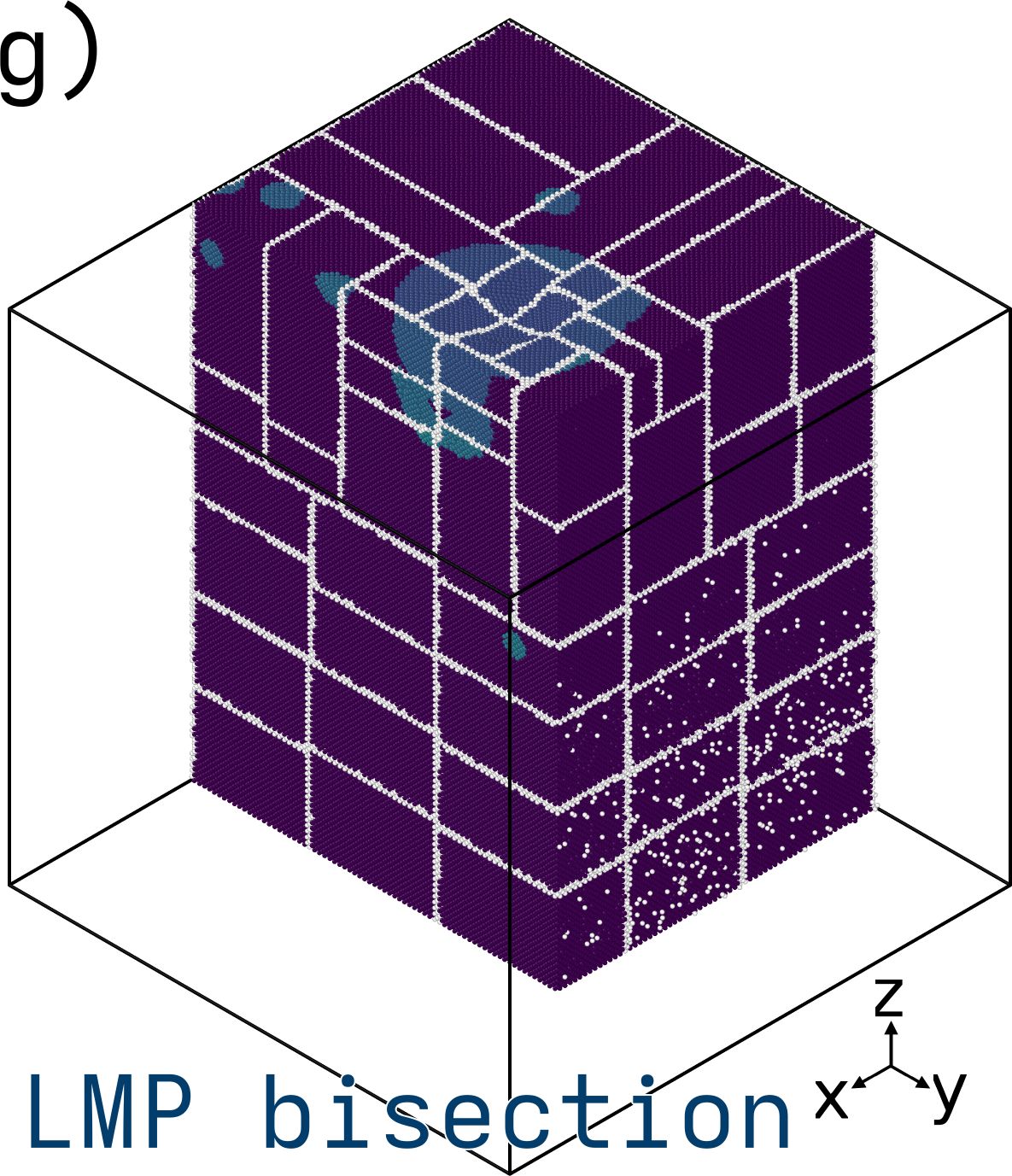}
\includegraphics[width=0.24\textwidth]{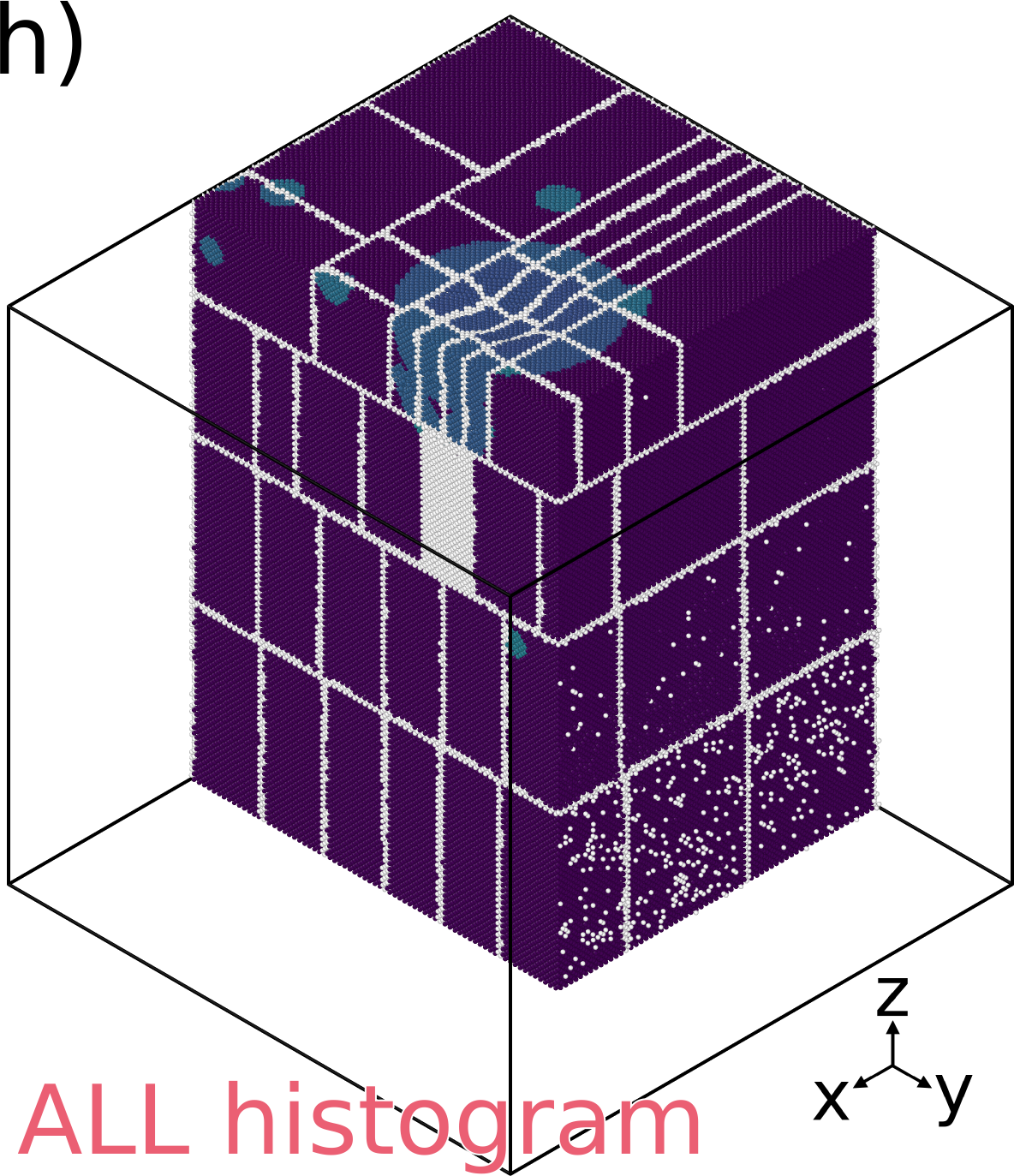}
\includegraphics[width=0.24\textwidth]{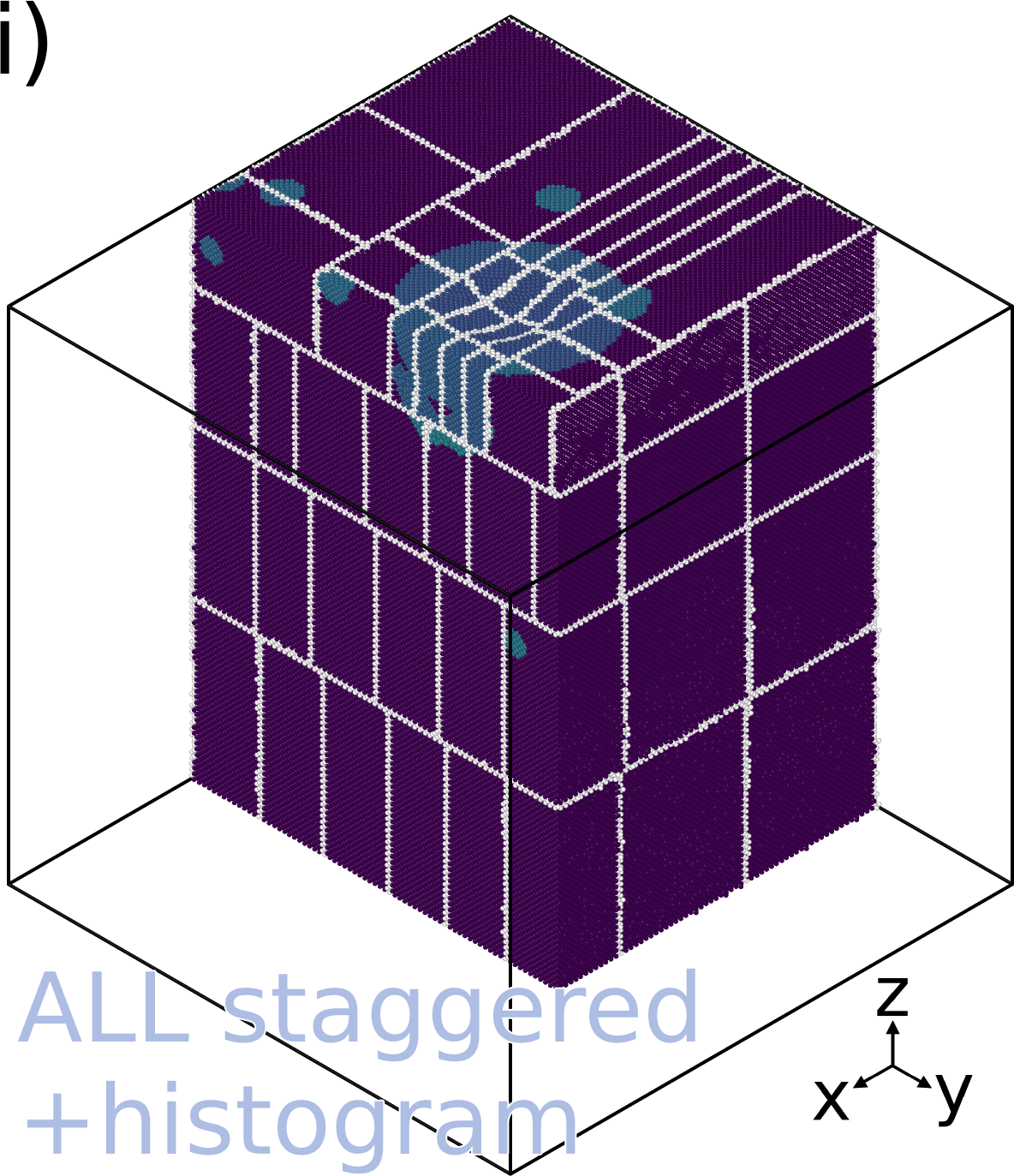}
\includegraphics[width=0.24\textwidth]{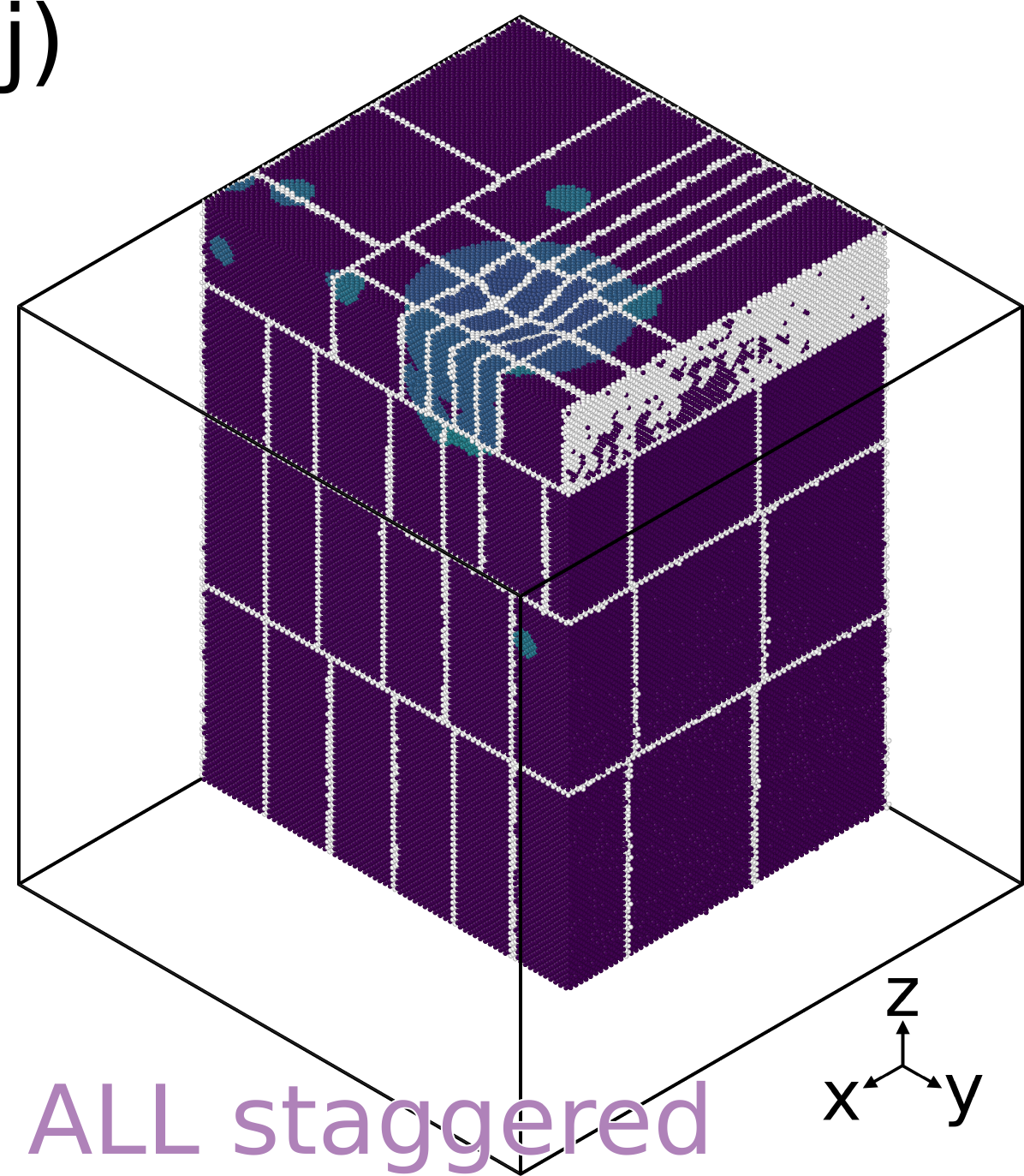}
\caption{\label{fig:nontensor}
Analysis of the load-balancing methods for staggered and irregular grids including
a) the total time spent in some parts of a simulation;
b) the wall-time and c) imbalance dependent on the timestep;
cumulated histograms measured at the end of the simulations of the distribution of the d) computational load, e) number of atoms and f) surface-by-volume ratio among the processors;
and g-j) visualizations of the domain decomposition after the final timestep of the simulations.
}
\end{figure*}

We compare the load-balancing methods for a staggered grid offered by the library ALL (cf. \cref{sec:load:balancing:methods}) with the irregular-grid load balancing of LAMMPS (cf. \cref{sec:load:balancing:methods}) in $\SI{1}{\pico\second}$-long simulations of the in \cref{sec:nanoindentation} introduced system.
Plots regarding the analysis are shown in \cref{fig:nontensor}.

\Cref{fig:nontensor}a shows the fraction of wall time spent for the whole simulation, communication, force calculation and load balancing.
On average 47\% of the time is spent for the force-calculation and 24\% for communication for the load-balancing methods (15\% and 49\% for the reference simulation without load balancing).
Thus, the fraction of wall-time spent for communication is far lower compared to the 45\% for regular grids (cf. \cref{fig:tensor}a).
Minimum and maximum communication duration vary by a factor between 1.4 (``histogram'') and 2.6 (``staggered'').
Some variation is expected as the surfaces of domains vary and there are no periodic boundary conditions in $z$.
Thus, the varying communication-times do not indicate significant idle times during communication.
The wall time spent for communication is minimal for the ``bisection'' method.
The force-calculation time is minimal for the ``bisection'' method and between 1.06 and 1.32 times larger for the other load-balancing methods and 4.15 times larger for the reference simulation without load balancing.

Measuring the time spent dependent on the timestep (cf. \cref{fig:nontensor}b) shows that the required wall time is similar for all but the method ``staggered''.
The imbalance measured dependent on the timestep (cf. \cref{fig:nontensor}c) reveals that the method ``staggered'' cannot adapt to the load distribution instantly but requires some iterations to reach an imbalance near 1.
This behavior is expected due to the local nature of this load-balancing method (cf. \cref{sec:load:balancing:methods}).
The other load-balancing methods reach a imbalance near 1 directly.
The average imbalance in the last 500 steps of the simulation is $\SI{1.056 \pm 0.016}{}$ for ``histogram'',  $\SI{1.062 \pm 0.017}{}$ for ``bisection'', $\SI{1.111 \pm 0.027}{}$ for ``histogram+staggered'', $\SI{1.148 \pm 0.040}{}$ for staggered and $\SI{6.054 \pm 0.027}{}$ for the reference simulation.

The cumulated histograms of the load per processor that were measured at the end of the simulation (cf. \cref{fig:nontensor}d) show visually that the methods ``bisection'' and ``histogram'' provide a balanced system.
The computational load of the ``histogram'' and ``bisection'' method differ for two reasons:
There are double computations to enable load balancing and the computation of the switching parameter strongly depend on the domain decomposition~\cite{adaptive_precision_potentials}.
The distribution of the number of atoms per domain is quite similar (cf. \cref{fig:nontensor}e), but the distribution of the surface-by-volume ratio shows significant differences (cf. \cref{fig:nontensor}f):
The largest and the average surface-by-volume ratio is smaller for the ``bisection'' method than for the three staggered-grid methods.
The surface-by-volume ratio is an indicator for the number of atoms near domain boundaries that need to be communicated at every step.
Thus, the surface-by-volume ratio explains why the total wall time required for communication is smaller for the ``bisection'' method than for the staggered-grid methods.
The domain boundaries at the end of the simulations are shown in \cref{fig:nontensor}g-j and help to understand the surface-by-volume histogram: The cuts $y$ through the top $z$ layer of the staggered grid produce quite narrow rows near the indentation point.
Thus, there are domains with large surface-by-volume ratios next to the indentation point in $\pm x$ direction.
In contrast, the domains of the ``bisection'' method look more like cubes that would have the ideal surface-by-volume ratio and would minimize communication.

\subsection{Ideal load-balancing frequency}
\label{sec:ideal:load:balancing:frequency}
The duration of one load-balancer call $\tau^\text{LB}$ is given by the duration $\tau^\text{DS}$ of shifting the domains.
LAMMPS rebuilds the neighbor list after every load-balancing~\cite[p. 1392]{lammps_manual}.
Therefore, the cost of of a load-balancer call is increased by the duration $\tau^\text{NL}$ of a neighbor list build, i.e.,
\begin{equation}
\tau^\text{LB} = \tau^\text{DS} + \tau^\text{NL}\,.
\label{eq:cost:load:balancing}
\end{equation}
Both contributions $\tau^\text{DS}$ and $\tau^\text{NL}$ can be measured within LAMMPS.

\begin{figure}[tb]
\includegraphics[width=2.7in]{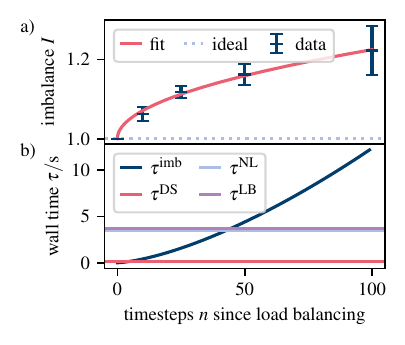}
\includegraphics[width=2.0in]{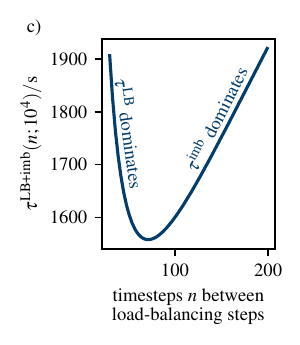}
\caption{\label{fig:lbevery}
a) Average imbalance after timestep 500 measured in four simulations with different load-balancing frequencies. The error bars denote the minimum and maximum measured imbalance.
b) Comparison of the cost of load balancing (cf. \cref{eq:cost:load:balancing}) and a load imbalance (cf. \cref{eq:gain:load:balancing}) dependent on the number of timesteps between load balancing.
c) Total cost of load balancing and imbalance (cf. \cref{eq:tau:lb:imb}) dependent on the load-balancing frequency.
}
\end{figure}

The load-balancer decreases the imbalance $I$ (cf. \cref{eq:imbalance}), ideally to 1.
Thereby, all processors $p$ should spent the same time $\tau^\text{F}$ for the force calculation in the next timestep.
However, as the simulation progresses some switching parameters $\lambda$ (cf. \cref{eq:gamma}) will change or atoms will migrate between domains.
Therefore, the imbalance increases with the number of timesteps $n$ since the last load-balancing step.
Under the assumption of slowly changing switching parameters, the total force calculation time ($\sum_p\tau_p^\text{F}$) is independent of $n$, but changes due to migrating atoms.
This assumption is valid for our test system after 500 timesteps as the imbalance fluctuates only slightly (cf. \cref{fig:nontensor}c).
We've measured the imbalance $I$ dependent on the number of timesteps since the last load-balancing by performing three additional simulations with less frequent RCB-load balancing.
Average and standard deviation of the imbalance after timestep 500 are shown in \cref{fig:lbevery}a.
The imbalance is found to be given by $I(n) = 1 + c \sqrt{n}$, where $c=0.022$ applies in this case, but the constant $c$ is expected to be strongly dependent on the system and the number of used processors $P$.
The approximation of $I(n)$ allows the calculation of the additional wall-time caused by load imbalance, i.e.,
\begin{equation}
\tau^\text{imb}(n) = \tau^\text{F} \sum_{n^\prime=0}^n \left(I(n) - 1\right)\,,
\label{eq:gain:load:balancing}
\end{equation}
where $\tau^\text{F}= 0.82\pm0.13\si{\second}$ was measured on average in the four simulations with different load-balancing frequencies.
\Cref{eq:gain:load:balancing} is compared with the average cost $\tau^\text{LB}$ (cf. \cref{eq:cost:load:balancing}) of a load-balancing step in \cref{fig:lbevery}b.
The combined cost of imbalance $\tau^\text{imb}$ and load-balancing $\tau^\text{LB}$ expected in a longer simulation of $N$ timesteps is given by
\begin{equation}
\tau^\text{LB+imb}(n;N) = \frac{N}{n} \left(\tau^\text{LB} + \tau^\text{imb}(n)\right)
\label{eq:tau:lb:imb}
\end{equation}
dependent on the number of timesteps $n$ between load balancing.
\Cref{fig:lbevery}c shows that ideally $n=71$ applies, which is significantly larger than the used $n=25$ in Ref.~\cite{adaptive_precision_potentials}.
Thus, one should dynamically balance a short simulation with the preferred load-balancing style and observe $I(n)$ in a second short simulation with \emph{LMP report} as load-balancing style.
Thereby, one can minimize the total cost of imbalance and load-balancing according to \cref{eq:tau:lb:imb}.

\section{Conclusion}
In the present work, we compared load-balancing methods for regular and staggered grids provided by the library ALL with the load balancing methods for regular and irregular grids already implemented in LAMMPS.
We applied these load-balancing methods for simulations that use an adaptive-precision EAM-ACE potential.
These simulations require load-balancing as the precise ACE potential is used only for a subset of atoms while the remaining system is simulated using the two orders of magnitude faster EAM potential.

Our results show that the ``tensor max`` method of ALL outperforms the ``shift'' method of LAMMPS: it reaches an imbalance that is 1.3 smaller (cf. \cref{fig:tensor}).
The advantage of the ``tensor max'' method is that it minimizes the maximum load of a processor between cuts of a regular grid.
Thereby, the load imbalance of the force computation is minimized.
The drawback of this method is that the communication volume is increased as this approach generates high surface-by-volume ratios.
Nevertheless, the increased communication volume is negligible compared to the minimized imbalance of the force computation.
In contrast, the ``shift'' method of LAMMPS balances the total load between two cuts of the regular grid domain decomposition.
Thereby, it generates better surface-by-volume ratios but a worse imbalance.
Therefore, LAMMPS benefits from the ``tensor max'' implementation that was developed for the present work.

The library ALL suggest a new domain decomposition for a regular grid based on the total work per domain.
Thus, ALL implicitly assumes a homogeneous load-distributions that is obviously not given for the given AP potential.
Therefore, the usage of the atomistic load by LAMMPS' ``shift'' method outperforms ALL's ``tensor classic'' method.
Note that the ``tensor max'' method outperforms the ``shift'' method of LAMMPS despite this wrong assumption.
Thus, LAMMPS would benefit even more from an implementation of the ``tensor max'' method that considers the atomic load instead of the load of a domain.

The comparison between load-balancing methods for staggered and irregular grids shows that both LAMMPS' ``bisection'' method and ALL's ``histogram'' method can provide a balanced system (cf. \cref{fig:nontensor}).
Calculating new domain boundaries and communicating all atoms to their new processor is significantly faster for a staggered grid than for an irregular grid.
The domain boundaries of a staggered grid can be shifted by ALL, while the domain boundaries of the irregular grid cannot be shifted.
Instead, the irregular domains need to be recomputed starting from the first bisection, which may result in two processors swapping their domains and causes the expensive migration of many atoms to new domains~\cite{lammps}.
Therefore, dynamic load balancing is faster for a staggered grid than for an irregular grid created by recursive bisections.
However, every new domain decomposition requires building a new neighbor list (cf. \cref{eq:cost:load:balancing}) and the cost of the load-balancing itself is negligible in comparison to the neighbor list build (cf. \cref{fig:lbevery}).

The advantage of the irregular grid of domains compared to the staggered grid is the significantly lower surface-by-volume ratio that is achieved through recursive bisections compared to the staggered grid of $P_\text{x}P_\text{y}P_\text{z}$ domains.
Therefore, the communication required for the time integration of the equations of motion is more expensive for a staggered grid than for the irregular grid (cf. \cref{fig:nontensor}a).
This communication cost difference is relevant at every timestep and makes the decreased load-balancing time of a staggered grid negligible for the system under study.
However, one should note that the communication cost for an adaptive-precision potential is larger than for constant-precision potentials that usually have a smaller force-cutoff.
Furthermore, the communication costs increase with the number of used processors.
Therefore, the current implementation of a staggered-gird domain decomposition is not beneficial for our adaptive-precision simulations with LAMMPS, but might be for simulations with constant precision potentials or fewer processors.
However, an updated implementation of a staggered grid that is more flexible and allows for an adjustable number of domains per row, an adjustable number of rows per layer and an adjustable number of layers could be promising.
Such a more dynamic staggered grid could improve the surface-by-volume ratio of domains and thus decrease the communication cost at every timestep.
Furthermore, one could improve the implementation of the communication routine of the staggered grid.
Currently it is implemented similarly to the communication of the irregular grid.
A future version could use the structure of the staggered grid to accelerate the communication like it is done by LAMMPS for the regular grid domain decomposition.

The analysis of the ideal load-balancing frequency (cf. \cref{sec:ideal:load:balancing:frequency}) has shown that load balancing every 71 timesteps would be ideal for the method ``histogram'' in our simulations.
In contrast, load balancing every 25 timesteps has decreased the overall computational cost compared to load balancing every 100 timesteps in a previous study~\cite{adaptive_precision_potentials}.
Therefore, our analysis suggests that the ideal load-balancing frequency strongly depends on the system under study.
Hence, one should run a short test simulation prior to a long production simulation as described in \cref{sec:ideal:load:balancing:frequency}: starting from a balanced system, one can measure the increase of the imbalance $I$ (cf. \cref{eq:imbalance}) in the absence of load-balancing to calculate the ideal load-balancing frequency.
The change of the imbalance will depend on the dynamics of the system, in particular of the migration of precisely calculated atoms.
Thus, the ideal load-balancing frequency will change when the dynamic of the system changes.

\begin{credits}
\subsubsection{\ackname}
The authors gratefully acknowledge the computing time on the supercomputer JURECA-DC~\cite{jureca} at Forschungszentrum Jülich under the project `slchem'.

\subsubsection{\discintname}
G. Sutmann is part of the program committee of the ``11th Workshop on Language-Based Parallel Programming Models'' at the PPAM 2026. D. Immel declares no competing interests.

\subsubsection{Code availability}
The LAMMPS code used for the load-balancing study is available at Ref. \cite{lammps_used}.

\end{credits}

\bibliographystyle{splncs04}
\bibliography{references.bib}

\end{document}